\documentclass[%
 reprint,
 amsmath,amssymb,
 aps,
]{revtex4-2}

\usepackage{graphicx}
\usepackage{dcolumn}
\usepackage{bm}
\usepackage{hyperref}
\usepackage[mathlines]{lineno}

\usepackage{graphicx,color}

\newcommand{\be}{\begin{equation}}
\newcommand{\ee}{\end{equation}}
\newcommand{\ba}{\begin{eqnarray}}
\newcommand{\ea}{\end{eqnarray}}

\newcommand{\lp}{\left(}
\newcommand{\rp}{\right)}

\newcommand{\w}{\wedge}

\begin{document}

\title{Classical scale-separated AdS$_3$ vacua in heterotic string theory}

\author{George Tringas}
 \email{georgios.tringas@lehigh.edu}
\author{Timm Wrase}%
 \email{timm.wrase@lehigh.edu}
\affiliation{%
Department of Physics, Lehigh University, \\
16 Memorial Drive East, Bethlehem, PA 18018, USA
}

\date{\today}

\begin{abstract}
We provide the first scale-separated AdS solutions from compactifications of the heterotic string. Our solutions have parametrically weak coupling, large volume, and the internal KK scale is parametrically smaller than the AdS length. These AdS$_3$ vacua preserve $\mathcal{N}=1$ supersymmetry and arise from compactifications on $G_2$ structure manifolds with $H$ flux and (smeared) gravitational instantons. All geometric moduli are stabilized, fluxes quantized, and the solutions are parametrically controlled in an appropriate large-flux limit.
\end{abstract}

\maketitle

\section{\label{sec:intro}Introduction}

The construction of scale-separated AdS vacua in string theory remains a central open problem \cite{Coudarchet:2023mfs}. In particular, finding controlled solutions that exhibit both parametric scale separation and weak coupling/large volume has proven difficult in ten-dimensional supergravity.
Recently, classes of three-dimensional AdS solutions of massive type IIA supergravity with these properties were obtained by compactifying on a toroidal $G_2$ holonomy orbifold with smeared O2/D2 and O6/D6 sources~\cite{Farakos:2020phe, VanHemelryck:2022ynr, Farakos:2023nms, Farakos:2025bwf}, while related constructions have appeared in type IIB supergravity  compactifications on $G_2$ structure spaces~\cite{VanHemelryck:2025qok}. 

As shown in \cite{Miao:2025rgf}, performing three T-dualities along internal torus directions interconnects these two approaches that lead to scale-separated vacua. Furthermore, it was shown in \cite{Emelin:2021gzx} that one can include O9-planes and D9-branes and thereby obtain classical AdS$_3$ solutions in type I string theory, in which all moduli are stabilized and fluxes are quantized. Such vacua point to an S-dual construction in the $SO(32)$ heterotic string theory, whose compactification on $G_2$ structure spaces was studied in \cite{deIaOssa:2019cci}. However, on the heterotic side the specific properties of AdS$_3$ vacua have not yet been worked out.

In this letter we show explicitly that parametrically scale-separated solutions can, in fact, be realized at weak heterotic coupling and large internal volume, thus providing the first fully classical, scale-separated AdS$_3$ vacua of the heterotic string theory.
The Yang–Mills gauge field in the adjoint representation of the heterotic gauge group will not play a significant role in our construction of scale-separated vacua, and these vacua therefore arise in both the $SO(32)$ and $E_8 \times E_8$ heterotic string theories.

\section{\label{sec:level2}Heterotic string compactifications on $G_2$ spaces}

We study compactifications of ten-dimensional heterotic string theory on a compact seven-manifold $X$ with $G_2$ structure.
The geometry, fluxes and gauge bundle form an $\mathcal{N}=1$ heterotic $G_2$ system~\cite{delaOssa:2017pqy}. 
Such systems are specified by the associative three-form $\Phi$ of $G_2$ spaces and the modified NSNS flux
\begin{equation}\label{eq:Hdefinition}
H = dB + \frac{\alpha'}{4}\!\left(\omega_{3,\mathrm{YM}}(A) - \omega_{3,\mathrm{L}}(\Theta)\right)\,,
\end{equation}
where $B$ is the Kalb-Ramond two-form field and $\omega_{3,\mathrm{YM}}(A)=\mathrm{Tr}\!\left(A\wedge dA+\frac{2}{3}\,A\wedge A\wedge A\right)$ is the Yang–Mills Chern–Simons three-form built from the gauge connection $A$. $\Theta$ is the spin connection on the tangent bundle, and the corresponding Lorentz Chern–Simons term, $\omega_{3,\mathrm{L}}(\Theta)$, is defined analogously.

Dimensional reducing the ten-dimensional heterotic supergravity action to three dimensions yields an $\mathcal{N}=1$ supergravity whose real superpotential is given by~\cite{deIaOssa:2019cci} (see also footnote \footnote{We use 10d Einstein frame that makes it easier to use S-duality between the heterotic string and the type I string theory to compare with results in \cite{Miao:2025rgf}. We also made explicit a dilaton factor $e^{2\phi}$ that was implicit or missing in the $H_7$ term in the superpotential in \cite{deIaOssa:2019cci}.} for a frame-related factor)
\begin{equation}\label{eq:P}
    P=\int \frac{e^{\frac{\phi}{2}} H_7 -e^{-\frac{\phi}{2}} \star\Phi \wedge H +\frac12 \Phi\wedge d\Phi }{4 \text{vol}(X)^2}  \,,
\end{equation}
where $\text{vol}(X)=\frac{1}{7}\int\Phi\wedge\star\Phi$ \footnote{We use $\star$ to denote the Hodge star on the internal $\mathrm{G}_2$ space and $\star_{10}$\ for the 10d Hodge star.}.
We see three different terms. The first arises from the spacetime filling Freund-Rubin $H_{\rm ext}$ flux, which we have written above as the internal flux $H_7 = e^{-\phi} \star_{10} H_{\rm ext}$.
The second term arises from the internal $H$ flux, while the last term is absent for $G_2$ holonomy manifolds, that have $d\Phi=0$, but is present for $G_2$ structure spaces with non-vanishing torsion.

Supersymmetric vacua are obtained by extremizing the superpotential with respect to the moduli. Varying $P$ with respect to the dilaton and volume modulus, we find
\begin{align}
    0=&\,\int e^{\frac{\phi}{2}} H_7 +\int e^{-\frac{\phi}{2}}\star\Phi\wedge H\,,\label{eq:partialdialton}\\
    0=&\, 7\int e^{\frac{\phi}{2}} H_7+2\int \Phi\wedge d\Phi - 5\int e^{-\frac{\phi}{2}} \star\Phi\wedge H \label{eq:partialvolP}\,,
\end{align}
where we used that $\Phi \sim {\rm vol}(X)^{\frac37}$ and $\star \Phi \sim {\rm vol}(X)^{\frac47}$.
The equations \eqref{eq:partialdialton}-\eqref{eq:partialvolP} combined imply that 
\begin{equation}\label{eq:genearlsolution}
\int e^{\frac{\phi}{2}} H_7 = -\int e^{-\frac\phi2} \star\Phi\wedge H=-\frac16 \int\Phi\wedge d\Phi \,.
\end{equation}
Thus, a non-trivial supersymmetric extremum requires all three types of terms to be present and nonzero. In that case we can use the above equation in $P$ and find that the value of the scalar potential at the minimum is
\begin{equation}\label{eq:Vmin}
V_{\rm min} =-4P^2 = - \frac{\left( \int e^{\frac{\phi}{2}} H_7\right)^2}{4 {\rm vol}(X)^4}\,.
\end{equation}
In string units, the integral of $H_7$ defines an integer-quantized flux number $h_7$ that we can make arbitrarily large. 
It is easy to read off from equation \eqref{eq:genearlsolution} that in this limit $e^\phi \propto (h_7)^{-\frac12}$ and ${\rm vol}(X)\propto (h_7)^{\frac78}$. 
This means in the large flux limit $h_7 \gg 1$, we find trustworthy supergravity solutions with large volume and at weak coupling, so $\alpha'$ and string loop corrections are highly suppressed.
However, as in all Freund-Rubin-type compactifications, using equation \eqref{eq:Vmin} we find that the AdS Hubble scale $L_{\text{AdS}} \sim 1/\sqrt{V_{\rm min}}$ scales in the same way as the KK-scale $L_{\text{KK}} \sim {\rm vol}(X)^{\frac87}$ \footnote{$L_{\rm KK}$ scales like the characteristic length of the internal space, i.e. $\mathrm{vol}(X)^{1/7}$. In going to 3D Einstein frame, an additional factor of $\mathrm{vol}(X)$ appears in $L_{\rm KK}$.}, namely $L_{\text{AdS}} \sim L_{\text{KK}} \sim h_7$.
This means that solutions in which \emph{only} the flux quantum $h_7$ is large cannot be described as lower-dimensional effective AdS$_3$ solutions, since the internal and external length scales are of the same order. 
They are only 10d solutions of the heterotic string theory on AdS$_3 \times X$. 
To remedy this general feature, we also have to make some of the internal $H$ flux quanta large.
Since we are working on a $G_2$ structure space whose associative three-form $\Phi$ is not closed, we must likewise allow for the internal $H$ flux to be non-closed, which induces charges on the compact space $X$ that must be canceled. We will discuss this in detail in the following sections.

\section{Compactifications on $\mathrm{G}_2$ spaces}

We consider a compact seven-dimensional $\mathrm{G}_2$ structure manifold $X$ with associative three-form $\Phi$ and its Hodge dual, the co-associative four-form $\Psi=*\Phi$.
The Betti numbers are $b^0=1$, $b^1\geq 0$, $b^2\geq 0$, $b^3>0$ and by Hodge duality $b^{7-p}=b^p$ \cite{joyce2000compact}.
We expand $\Phi = s^i \Phi_i$, where the $\Phi_i$, ($i=1,\ldots,b^3$) are a basis of $H^3(X,\mathbb{Z})$. The real coefficients $s^i$ are the geometric moduli that we have to stabilize.
In addition, there is the dilaton $e^\phi$ and potentially further moduli arising from the Kalb-Ramond two-form $B$ and/or the one-form gauge field $A$.

For $G_2$ structure manifolds the associative and co-associative forms need not be closed or co-closed, and their exterior derivatives decompose into $G_2$ irreducible representations 
\begin{align}
d \Phi &= \tau_0 \star \Phi + 3 \tau_1 \w \Phi + \star \tau_3\,,\label{eq:dPhi}\\
d \star \Phi &= 4 \tau_1 \w \star \Phi + \star \tau_2\,,\label{eq:dstarPhi}
\end{align}
where the four torsion classes $\tau_i$ are $i$-forms, and lie in the \textbf{1}, \textbf{7}, \textbf{14}, and \textbf{27} representations of the $G_2$ structure group, respectively. 

Varying the superpotential $P$ in equation \eqref{eq:P} with respect to the connections $A$ and $\Theta$ (cf. equation \eqref{eq:Hdefinition}) yields the $\mathrm{G}_2$ instanton conditions
\begin{equation}\label{eq:instantons}
F\!\wedge\!\star\Phi = 0\,, 
\qquad 
R\!\wedge\!\star\Phi = 0\,.
\end{equation}
Varying with respect to $B$ we find 
\begin{equation}\label{eq:partial_BofP}
    d\left(e^{-\frac{\phi}{2}}\star\Phi\right)=0\quad\Rightarrow
    \quad d\star\Phi=\frac{1}{2}d\phi\wedge\star\Phi\,,
\end{equation}
which, using \eqref{eq:dstarPhi}, implies $8\tau_1=d\phi, \tau_2=0$. For $\tau_2 = 0$, the $\mathrm{G}_2$ structure is integrable, and the $G_2$ metric connection has totally antisymmetric torsion $T$, which satisfies $T = H$, see \cite{delaOssa:2016ivz}.
To compute $\partial_{s^i} P$, we use \cite{Grigorian:2008tc}
\begin{equation}
\partial_{s^i} (\star \Phi) = \frac73 \frac{\int \Phi_i \w \star \Phi}{\int \Phi \w \star \Phi} \, \star \Phi - \star \Phi_i\,.
\end{equation}
This then leads to 
\begin{eqnarray}
\partial_{s^i} P &=& \frac{ \int \Phi_i \w \star \Phi}{6 {\rm vol}(X)^3}  \left( -\int e^{\frac{\phi}{2}} H_7 +\frac12 \int e^{-\frac{\phi}{2}}  \star \Phi \w H \right. \cr
&&\qquad\qquad\quad\left.-\frac12 \int \Phi \w d\Phi \right) \\
& +&\frac{1}{4 {\rm vol}(X)^2} \lp \int e^{-\frac{\phi}{2}}  \star \Phi_i \w H + \int \Phi_i \w d\Phi \rp\,.\nonumber
\end{eqnarray}
Since $ \int  e^{-\frac{\phi}{2}} \star \Phi_i \w H= \int  e^{-\frac{\phi}{2}} \Phi_i \w  \star H$, and the $\Phi_i$ are a basis of left-invariant three-forms, we find, using \eqref{eq:genearlsolution}, that \mbox{$\partial_{s^i} P=0$} reduces (after acting with $\star$) to
\begin{equation}\label{eq:Phi}
\Phi = -\frac{{\rm vol}(X)}{\int e^{\frac{\phi}{2}} H_7} \lp e^{-\frac{\phi}{2}} H + \star d\Phi \rp\,.
\end{equation}
This shows that $\Phi=s^i \Phi_i$ is generically fixed, and therefore all the geometric moduli $s^i$ are fixed by the above equation coming from $\partial_{s^i} P=0$.

Lastly, the $H$ flux satisfies the Bianchi identity 
\begin{equation}\label{eq:dH}
    dH = \frac{\alpha'}{4}\!\left(\mathrm{Tr}\,F\!\wedge\!F - \mathrm{Tr}\,R\!\wedge\!R \right) \,,
\end{equation}
where we included NS5-brane sources in the gauge instanton piece $\mathrm{Tr}\,F\!\wedge\!F$ \cite{Witten:1995gx}.

When solving the above equations to find scale-separated vacua, we restrict to $G_2$ structure spaces with $b^1=b^2=0$. We expand everything in left-invariant forms, which is expected to be a consistent truncation~\cite{Cassani:2009ck}, but not necessarily a low-energy effective action.
In our concrete example below we will show that it does define indeed a genuine three-dimensional low-energy theory.
To cancel internal charges we use localized instantons, but we can solve the equations only in the smeared limit (see section 4.1 in \cite{Junghans:2023lpo} for a related discussion of smeared O-planes).

For compact $G_2$ structure manifolds with $b^1=b^2=0$, we find that there are no moduli from expanding $A$ and $B$ along the internal space \footnote{The gauge field $A$ in 3d is dual to a scalar field that we do not stabilize.}.
We also have \mbox{$F=R=0,$} which trivially satisfies \eqref{eq:instantons}, and \eqref{eq:partial_BofP} is likewise automatically satisfied.

Choosing a basis of four-forms $\Psi^i\in H^4(X,\mathbb{Z})$ such that $\int \Phi_i \w \Psi^j = \delta_i^j$, we define
\begin{equation}
d \Phi_i = M_{ij} \Psi^j\,,
\end{equation}
where the matrix $M_{ij}$ is symmetric.
Its entries are related to the torsion classes above and are often called metric fluxes, since they can arise from T-dualizing $H$ flux.
This $H$ flux can also be present independently of the metric fluxes, and we expand it as $H = h^i \Phi_i$, where in string units $h^i \in \mathbb{Z}$.
Since we are compactifying on a seven-dimensional space, there are three external directions and $H$ can also have a component along them.
As mentioned above, this is dual to an $H_7$ flux proportional to the internal volume form and satisfying (in string units) $\int H_7 = h_7 \in \mathbb{Z}$.  

Since $dH = h^i M_{ij} \Psi^j$ is generically non-zero, the Bianchi identity \eqref{eq:dH} imposes constraints on the linear combinations $h^i M_{ij}$.
In general, $b^4$ linear combinations of the $b^3$ fluxes $h^i$ appear. Since $b^3=b^4$, all $h^i$ are generically fixed by the right-hand side of the Bianchi identity. 
We see below, that this would prevent us from taking any of the $h^i$ parametrically large.
It is crucial for our scale-separated AdS$_3$ vacua that this does not happen and that some of the $h^i M_{ij}$ are zero or linearly dependent.

\section{A concrete example}
As a concrete example, we focus on the toroidal orbifold $X = T^7 / \Gamma$, where the orbifold group $\Gamma=\mathbb{Z}_2^3$ has three generators $\Gamma =\{\Theta_{\alpha},\Theta_{\beta},\Theta_{\gamma}\}$ that act on the toroidal coordinates as follows
\begin{align}\label{Z2s}
\Theta_\alpha : y^a & \to (-y^1, -y^2, -y^3, -y^4, +y^5, +y^6, +y^7) \, , \nonumber
\\
\Theta_\beta : y^a & \to (-y^1, -y^2, +y^3, +y^4, -y^5, -y^6, +y^7) \, ,
\\
\Theta_\gamma : y^a & \to (-y^1, +y^2, -y^3, +y^4, \frac12 -y^5, +y^6, -y^7) \, . \nonumber
\end{align}
This toroidal orbifold can be resolved to a $G_2$ manifold with $b^2=0$ and $b^3=215$ \cite{joyce2000compact}. However, here we work at the orbifold point and restrict to the $\Gamma$-invariant three-forms descending from the torus, which gives $b^3=7$, and hence seven bulk moduli~$s^i$. 

We follow the conventions of \cite{Miao:2025rgf} for an explicit basis of the three-forms $\Phi_i$ and their dual 4-forms $\Psi^i$ in terms of the torus one-forms $dy^a$.
In this concrete example, for any fixed $i$, 
\begin{equation}\label{eq:starPsii}
    \star \Psi^i= \frac{(s^i)^2}{{\rm vol}(X)} \Phi_i\,,
\end{equation} 
and the volume is
\begin{equation}\label{eq:volume}
    {\rm vol}(X) = \frac17 \int \Phi \w \star \Phi = \left(s^1 s^2 \ldots s^7\right)^{\frac13}\,.
\end{equation}
To include non-vanishing curvature for our $G_2$ structure space, we generalize from the toroidal orbifold to a simple nilmanifold by replacing $dy^a \to e^a$.
The one-forms $e^a$ satisfy the structure equations
\begin{equation}
    de^a=\frac{1}{2}f^{a}_{bc}e^{b}\wedge e^{c}\,.
\end{equation}
We take $f^7_{56} \equiv f \neq 0$, with all other metric fluxes $f^a_{bc}=0$, corresponding to a class of nilmanifolds compatible with the orbifold we consider \cite{VanHemelryck:2025qok}. 
In the conventions of \cite{Miao:2025rgf} this implies $M_{12} = M_{21} = -f$, with all other $M_{ij}$ vanishing.
For the $H$ flux we make the ansatz 
\begin{equation}
    H=h \sum_{i=1}^2\Phi_i -N \sum_{j=3}^7\Phi_j\,. 
\end{equation}
The model is invariant under the exchange of $i=1$ and $i=2$, so that $s^1=s^2$, and likewise $s^3=s^4=s^5=s^6=s^7$.
These are stabilized by equation \eqref{eq:Phi},
\begin{equation}
\begin{split}
    s^1&=-\frac{\left(s^1\right)^{\frac23}\left(s^3\right)^{\frac53}}{e^{\frac{\phi}{2}} h_7} \lp e^{-\frac{\phi}{2}} h - \frac{\left(s^1\right)^{\frac73} f}{\left(s^3\right)^{\frac53}} \rp\,,\\
    s^3&=\frac{\left(s^1\right)^{\frac23}\left(s^3\right)^{\frac53}}{e^{\phi} h_7} N\,,
\end{split}
\end{equation}
where we used equations \eqref{eq:starPsii} and \eqref{eq:volume}.
The dilaton is stabilized by \eqref{eq:genearlsolution} as 
\begin{equation}
    e^{\frac{\phi}{2}} = \frac{\left(s^1\right)^2 f}{3 h_7} \,.
\end{equation}
Solving these three equations, we find that the fields are stabilized in terms of the fluxes as
\begin{align}
    e^\phi &= \frac{6 N^{\frac52}}{\left(h_7\right)^{\frac12} f h}\,,\quad\quad  \label{eq:solutiondilaton} \\
    s^1&=s^2 = \frac{54^{\frac14} \left(h_7\right)^{\frac38} N^{\frac{5}{8}}}{h^{\frac14} f^{\frac34}}\,,\\
    s^i&=\frac{2\times 54^{\frac14} \left(h_7\right)^{\frac38} N^{\frac{13}{8}}}{h^{\frac54} f^{\frac34}}\,\quad\text{for}\quad i=3,\dots,7\,.
\end{align}
To be in a parametrically controlled supergravity regime with suppressed string-loop and $\alpha'$ corrections, we impose $e^\phi\ll1$, perform a Weyl transformation to string frame $s^i = e^{-\frac34 \phi} s^i_s$, and require $s^i_s\gg1$.
In addition, we require the Kaluza–Klein scale to decouple from the AdS radius, i.e., $L_{\mathrm{AdS}}\gg L_{{KK}}$.
Note that there are several KK-towers, associated with the dilaton, and the seven $s^i$.
Moreover, on $G_2$ structure spaces it is not guaranteed that these are the lightest modes, and one should in principle compute the spectrum of the Laplacian. This was done for several examples in \cite{Andriot:2018tmb}, where it was found for simple nilmanifolds, similar to  ours, that these are indeed still the lightest modes. Instead of repeating such a calculation, we perform an alternative stringent check and show that, in an appropriate limit, all one-cycles of the underlying torus become parametrically large.

Since we do not analyze the full Kaluza–Klein spectrum, we will approximate the KK scale by the individual torus string-frame radii,
\begin{equation}
    r_{s,\{1,2,3,4\}}=\sqrt{r_{s,\{5,6\}}}=\sqrt{\frac{2N}{h}}r_{s,7}\,,\quad
    r_{s,7}=\sqrt{\frac{3N}{f}}\,.
\end{equation}
So, all torus radii in string frame become parametrically large if we take the unbounded flux quantum $N\gg 1$.

Computing the vacuum expectation value and the Kaluza–Klein scale for our solution, we find
\begin{align}
    V_{\text{min}}&=-\frac{1}{864^2}\frac{f^6h^8}{(h_7)^2N^{10}}\,,\\
    L_{\text{KK},i}^2&\sim \frac{432(h_7)^2N^5}{f^3h^4}\, r_{s,i}^2\,.
\end{align}
In the large $N\gg1$ limit, the largest internal circles are $ r_{s,5}= r_{s,6}$. Even for them we find a parametric separation of scales,
\begin{equation}
    \frac{L_{\text{KK}}}{L_{\text{AdS}}}\sim \frac{1}{N}\,.
\end{equation}
Thus, the relevant KK length scales are all parametrically separated from the AdS radius in the large $N\gg1$ limit.
Requiring weak coupling, large string-frame three-cycles, and parametric scale separation, we obtain the condition
\begin{equation}
    1 \ll N^5 \ll h_7 \,.
\end{equation}

Interestingly, for the simple setup above the masses squared of the eight scalar fields are integers in AdS units,
\be
\frac{m^2}{|V_{\rm min}|} = \{ 8,8,8,8,8,8,8,120\}\,.
\ee
This implies that the dual operators $\Delta$ in the CFT$_2$ have integer conformal dimensions, 
\begin{equation}
    \Delta\left(\Delta-2\right)=m^2L^2_{\text{AdS}}\,,\quad\quad \Delta=\{4,12\} \,.
\end{equation}
The appearance of integer conformal dimensions was observed in related solutions in \cite{Apers:2022tfm,Arboleya:2024vnp,VanHemelryck:2025qok,Proust:2025vmv}, and it is related to the method of tadpole cancellation \cite{Farakos:2025bwf,Miao:2025rgf}.

Lastly, we have to satisfy the tadpole cancellation condition
\begin{align}\label{eq:Bianchiexplicit}
    dH &= h M_{12} (\Psi^1 +\Psi^2)= -h f (\Psi^1 +\Psi^2)\cr
    &=\frac{\alpha'}{4}\!\left(\mathrm{Tr}\,F\!\wedge\!F - \mathrm{Tr}\,R\!\wedge\!R\right) \,.
\end{align}
For our $T^7/\mathbb{Z}_2^3$ orbifold we have a total of seven different non-trivial $\mathbb{Z}_2$ actions on the $T^7$.
Each of them leads to 16 singularities, locally of the form $\mathbb{R}^3 \times \mathbb{R}^4/\mathbb{Z}_2$, where the $\mathbb{Z}_2$ flips the sign of the four coordinates.
This latter part $\mathbb{R}^4/\mathbb{Z}_2$ is the singular limit of the Eguchi-Hanson space \cite{Eguchi:1978xp, Eguchi:1979yx}, where a $\mathbb{P}^1$ has shrunk to zero size at the singularity.
This means that at each singularity we have delta function source from $\mathrm{Tr}\,R\!\wedge\!R$.
With current techniques, we cannot solve the equations for intersecting sources even in flat space \cite{Bardzell:2024anh}.
We therefore resort to a smearing of these sources, replacing each of the delta functions by a constant (e.g. solving only the integrated Bianchi identity \eqref{eq:Bianchiexplicit}).
We then cancel the gravitational instanton $\mathrm{Tr}\,R\!\wedge\!R$ in the $\Psi^1+\Psi^2$ direction by fixing the product of the so far unspecified $h$ and $f$ flux.
Note, that the sign of their charge is fixed, since the dilaton solution in \eqref{eq:solutiondilaton} requires $h f>0$, which matches the sign of the gravitational instantons.
For the remaining charges in the other directions, we turn on gauge instantons that cancel the contributions from the gravitational instantons on our $G_2$ structure space.
For the heterotic $SO(32)$ string theory that is S-dual to type I string theory, this translates into using O5-planes and D5-branes, which are S-dual to the gravitational and gauge instantons, respectively.
Thus, the negative contribution of the O-planes in type II/I compactifications that is required to generate scale separation \cite{Gautason:2015tig,Tringas:2025uyg} is replaced by gravitational instantons in the equations of motion and scalar potential.

\section{Conclusion}
Arguably the most important goal in string phenomenology is moduli stabilization, i.e., giving masses to all the light scalar fields that usually arise in string compactifications.
There are very few explicit constructions that are broadly accepted, and none of these give rise to an effectively lower-dimensional theory.
The issue is to achieve scale separation so that the internal space is much smaller (i.e., with larger KK scale) than the external space.
For AdS compactifications, it was conjectured that scale separation is impossible \cite{Lust:2019zwm}, and this was supported for Einstein-space setups \cite{Bedroya:2025ltj} using the brane picture of AdS flux vacua \cite{Apers:2025pon}.
On the other hand, progress toward understanding scale-separated constructions related to \cite{DeWolfe:2005uu} has been made via uplifts to M-theory \cite{Cribiori:2021djm,VanHemelryck:2024bas} and through analyses of backreaction \cite{Junghans:2020acz,Marchesano:2020qvg,Emelin:2022cac,Junghans:2023yue}, which enhances their consistency.
The emergence of scale-separated AdS$_3$ vacua in various supergravities \cite{Farakos:2020phe,VanHemelryck:2022ynr,Farakos:2023nms,Farakos:2025bwf,VanHemelryck:2025qok,Miao:2025rgf} indicates that such vacua can be realized broadly \cite{Tringas:2025uyg}.

Here, we describe the first scale-separated AdS vacua for heterotic string theory.
Using standard ingredients, like fluxes and curvature, we show that a particular large flux limit gives parametrically weak coupling, large volume and scale-separation. 
The only immediately apparent issue with our solutions is that we cannot explicitly solve the non-linear Einstein equations with intersecting delta-function localized gravitational instantons. 
This decade old problem is commonly addressed by showing the existence of solutions in the smeared limit \footnote{O3-planes in type IIB/F-theory do not intersect. So, solutions like the ones in \cite{Demirtas:2021nlu} can circumvent this issue.}
Thus, our solutions are on equal footing with existing ones in type II string theory, but they may allow future refinements: since we only use the NSNS $H$ flux and a twisted toroidal orbifold (i.e., a simple nilmanifold), our constructions may admit a relatively simple worldsheet description.
Torsional heterotic compactifications have long been a subject of interest \cite{Strominger:1986uh,Hull:1986kz,Gauntlett:2003cy,Becker:2009df}, including their worldsheet sigma-model description \cite{Quigley:2011pv}.
Compactifications with $G_2$ holonomy have been studied in \cite{Font:2010sj,Acharya:2021rvh}.
In particular, the $G_2$ linear sigma model corresponding to~\cite{delaOssa:2017pqy}, which includes our setup, was studied in \cite{Fiset:2017auc} from the worldsheet point of view to explore the infinitesimal moduli of the heterotic string on $G_2$ structure geometries.
In addition, it might be useful in type II constructions to check whether the NSNS sector alone can, in certain cases, lead to a full moduli stabilization, which would then allow for a similar string worldsheet description.

\begin{acknowledgments}
We thank Maxim Emelin, Fotis Farakos, Zheng Miao, and Muthusamy Rajaguru for collaborating on closely related papers that inspired this work. We also thank Nicole Righi for discussions. This work is supported in part by the NSF grant PHY-2210271 and the Lehigh University CORE grant with grant ID COREAWD40.
\end{acknowledgments}


\bibliography{refs}

\end{document}